\documentclass{webofc}
\usepackage[varg]{txfonts}   
\begin{document}
\title{Creating a content delivery network for general science on the  internet backbone using XCaches}
%
%

\author{\firstname{Edgar}
\lastname{Fajardo}\inst{1}\fnsep\thanks{\email{emfajard@ucsd.edu}} \and
        \firstname{Derek} \lastname{Weitzel}\inst{2}\thanks{\email{dweitzel@unl.edu}}\fnsep \and
        \firstname{Mats} \lastname{Rynge}\inst{4}\fnsep\thanks{\email{rynge@isi.edu}} \and
        \firstname{Marian} \lastname{Zvada}\inst{2}\fnsep\thanks{\email{marian.zvada@cern.ch}} \and
        \firstname{John} \lastname{Hicks}\inst{5}\fnsep\thanks{\email{jhicks@internet2.edu}} \and
        \firstname{Mat} \lastname{Selmeci}\inst{6}\fnsep\thanks{\email{matyas@cs.wisc.edu}} \and
        \firstname{Brian} \lastname{Lin}\inst{6}\fnsep\thanks{\email{blin@cs.wisc.edu}} \and
        \firstname{Pascal} \lastname{Paschos}\inst{3}\fnsep\thanks{\email{paschos@uchicago.edu}} \and
        \firstname{Brian} \lastname{Bockelman}\inst{8}\fnsep\thanks{\email{BBockelman@morgridge.org}} \and
        \firstname{Andrew} \lastname{Hanushevsky}\inst{5}\fnsep\thanks{\email{abh@slac.stanford.edu}}\and
        \firstname{Frank} \lastname{W\"urthwein}\inst{1}\fnsep\thanks{\email{fkw@ucsd.edu}}
        \firstname{Igor} \lastname{Sfiligoi}\inst{1}\fnsep\thanks{\email{isfiligoi@sdsc.edu}}
}

\institute{University of California San Diego, 9500 Gilman Dr, La Jolla, CA 92093
\and
           University of Nebraska, 1400 R Street, Lincoln, NE 68588
           \and 
           University of Chicago, 5801 S Ellis Ave, Chicago, IL 60637
           \and
           University of Southern California, Information Sciences Institute, 4676 Admiralty Way, Suite 1001, Marina del Rey, CA 90292
           \and 
           SLAC National Accelerator Laboratory, 2575 Sand Hill Rd, Menlo Park, CA 94025
           \and
           Internet2, 100 Phoenix Drive, Suite 111 Ann Arbor, MI 48108
           \and 
           UW Madison, 1210 W. Dayton St, Madison, WI 53706
            \and 
           Mortgridge Institute, 330 N Orchard St, Madison, WI 53715
          }

\abstract{%
  A general problem faced by opportunistic users computing on the grid is that delivering cycles is simpler than delivering data to those cycles. In this project XRootD caches are placed on the internet backbone to create a content delivery network. Scientific workflows in the domains of high energy physics, gravitational waves, and others profit from this delivery network to increases CPU efficiency while decreasing network bandwidth use.
}
\maketitle
\section{Introduction}
\label{intro}
Scientific organizations face the problem of delivering data to computing workflows at a combination of owned and opportunistic clusters (sites). Since either the data is read locally at the organization owned sites (no use of opportunistic resources) or the data is accessed remotely with the drawback of high bandwidth utilization and performance degradation of the organization's storage systems (origins).
 
A third possible solution is the use of caches. Data accessed through the caches travels the network once, from the storage system to the caches. Once cached it can be delivered to multiple jobs thus using the network bandwidth efficiently while acting as a protection layer to the storage systems.  This caching strategy implies solving which caching technology to use (Section \ref{sec-stashcache}), how jobs access the caches (Section \ref{sec-cvmfs}), where to place them (Section \ref{sec-deployment}), how to deploy them (Section \ref{sec-kubernetes}) and analyzing their performance (Section \ref{sec-performance}). 

\section{Stashcache Background}
\label{sec-stashcache}
The Stashcache\cite{stashcache} service is a file block caching technology based on XRootD's \cite{dorigo2005xrootd} caching service XCache\cite{xcache}. XRootD's architecture is a tree-based structure of servers (bottom leafs) and redirectors (top leafs). Once a client requests a file from the redirector, it queries the servers below it. If they have the file the client is redirected to the server. If not the redirector contacts the redirector above it and so on. This type of tree structure is called a federation.

\begin{figure}[h]
\centering
\includegraphics[width=8cm,clip]{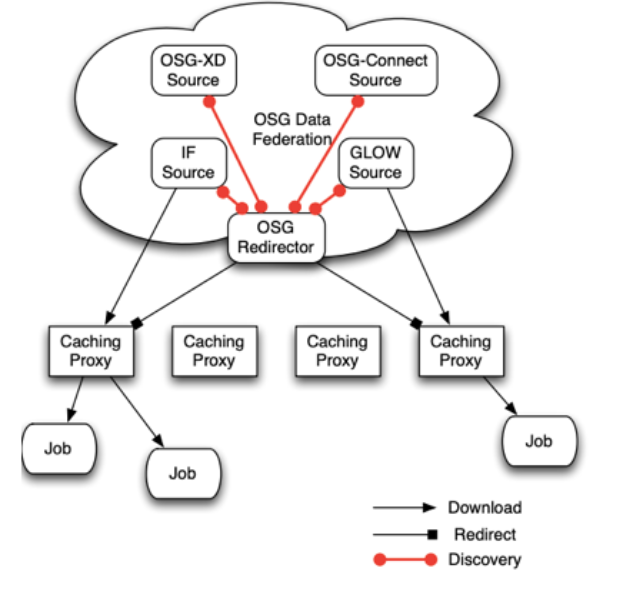}
\caption{Stashcache architecture.}
\label{stash-arch}       
\end{figure}

Stashcache sits in front of a federation. When a client requests a file to the cache, the cache will serve if it is already on  disk (cache hit). If not, the cache will contact the federation and retrieve the file from it and then serve it to the client (cache miss).  Stashcache has interfaces for the XRootD (propietary) and HTTP(s) protocols and its architecture be seen in Figure \ref{stash-arch}.

\section{Cache client access via CVMFS}
\label{sec-cvmfs}

The Stashcache architecture does not mention how users pick which cache to use. The CERN Virtual Machine File System (CVMFS)\cite{CVMFS} system solves this  problem using GeoIP. It also presents the users a POSIX-like file mount. When a client tries to open a file, CVMFS determines what is the closest geographical Stashcache and retrieves the file from it using HTTP. If the closest cache is unavailable or slowly responding then CVMFS tries the second closest and so on until it retrieves the file. Although CVMFS had been commonly used to distribute scientific software and small files (less than 100 MB experiment calibration data), the approach to distribute "experiment data" (files over 1 GB) was originally introduced in Ref \cite{cvmfsData}.

\subsection{Difference between Stashcache and Frontier squid}
Frontier Squid \cite{squid} is a widely deployed HTTP-based file caching technology used by CMVFS. There are two major differences between frontier squid and stashcache. A file in a squid cache has a limited lifetime: if the file gets changed in the stratum server (the squid equivalent of an origin),  after a few minutes the squid cache will pick up the change and serve the new version. If a file in a stashcache is changed on the origin, that change will not be picked up by stashcache and the cache will serve the old version of the file (until the file is purged from the disk). In other words, the squid model is "write few, read many," whereas the stashcache model is "write once, read many."

The second difference is in the size of files that can be served. Squid caches are optimized for handling small files O(MB) while Stashcache is meant for large files O(GB). Moreover, CVMFS is configured to have compute node level caching of files. This feature is not used by Stashcache however it is used by frontier squid. Files served by Stashcache are of the same size of the local partition hence one file filling the local partition can render useless the compute node level caching.

\section{Stashcache Deployment}
\label{sec-deployment}
\begin{figure}[h]
\centering
\sidecaption
\includegraphics[width=6cm,clip]{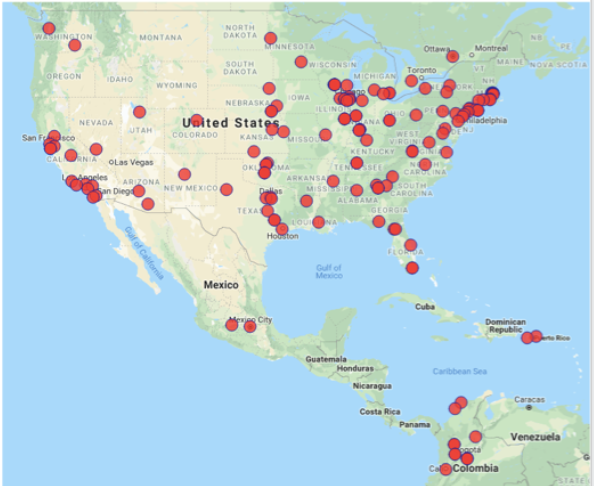}
\caption{Geographical location of OSG computing resources.}
\label{osgCOmputingResources}       
\end{figure}

\begin{figure}[h]
\centering
\sidecaption
\includegraphics[width=6cm,clip]{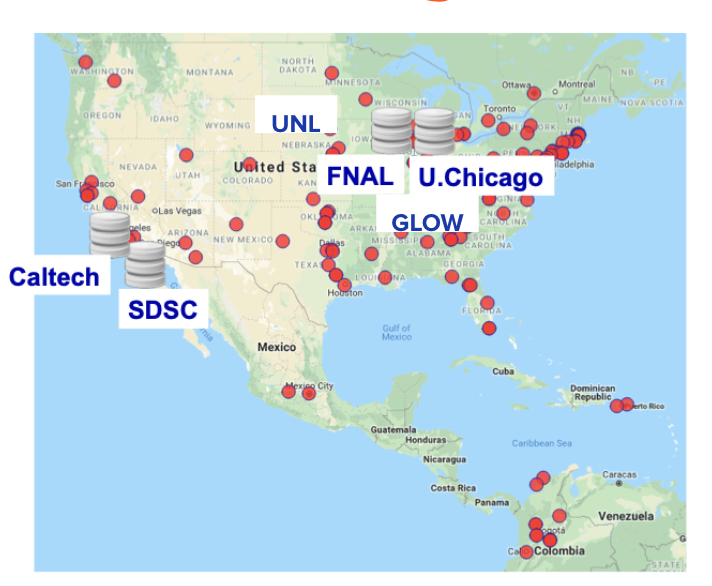}
\caption{Geographical location of data origins.}
\label{osgDataOrigins}       
\end{figure}

\begin{figure}[h]
\centering
\includegraphics[width=10cm,clip]{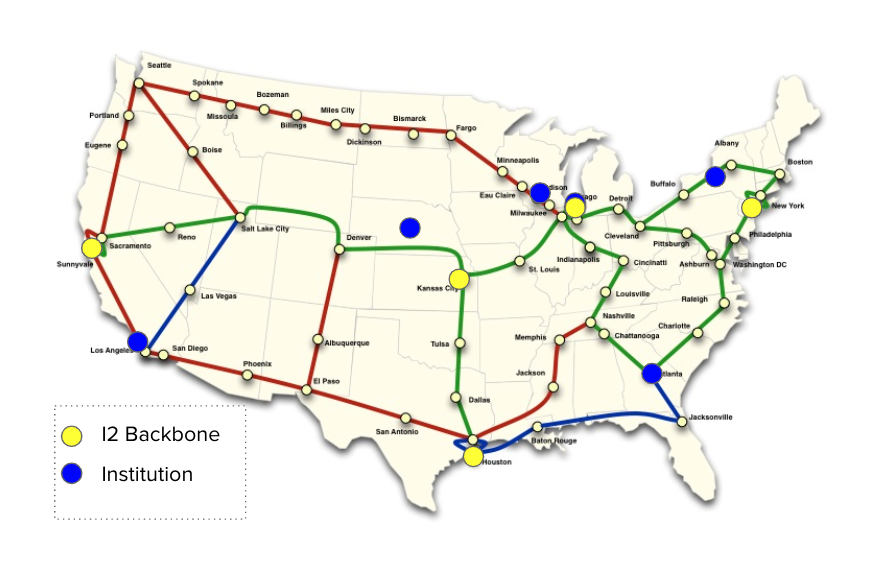}
\caption{Geographical location of Stashcache in the US internet backbone.}
\label{StashcacheUS}       
\end{figure}

\begin{figure}[h]
\centering
\includegraphics[width=10cm,clip]{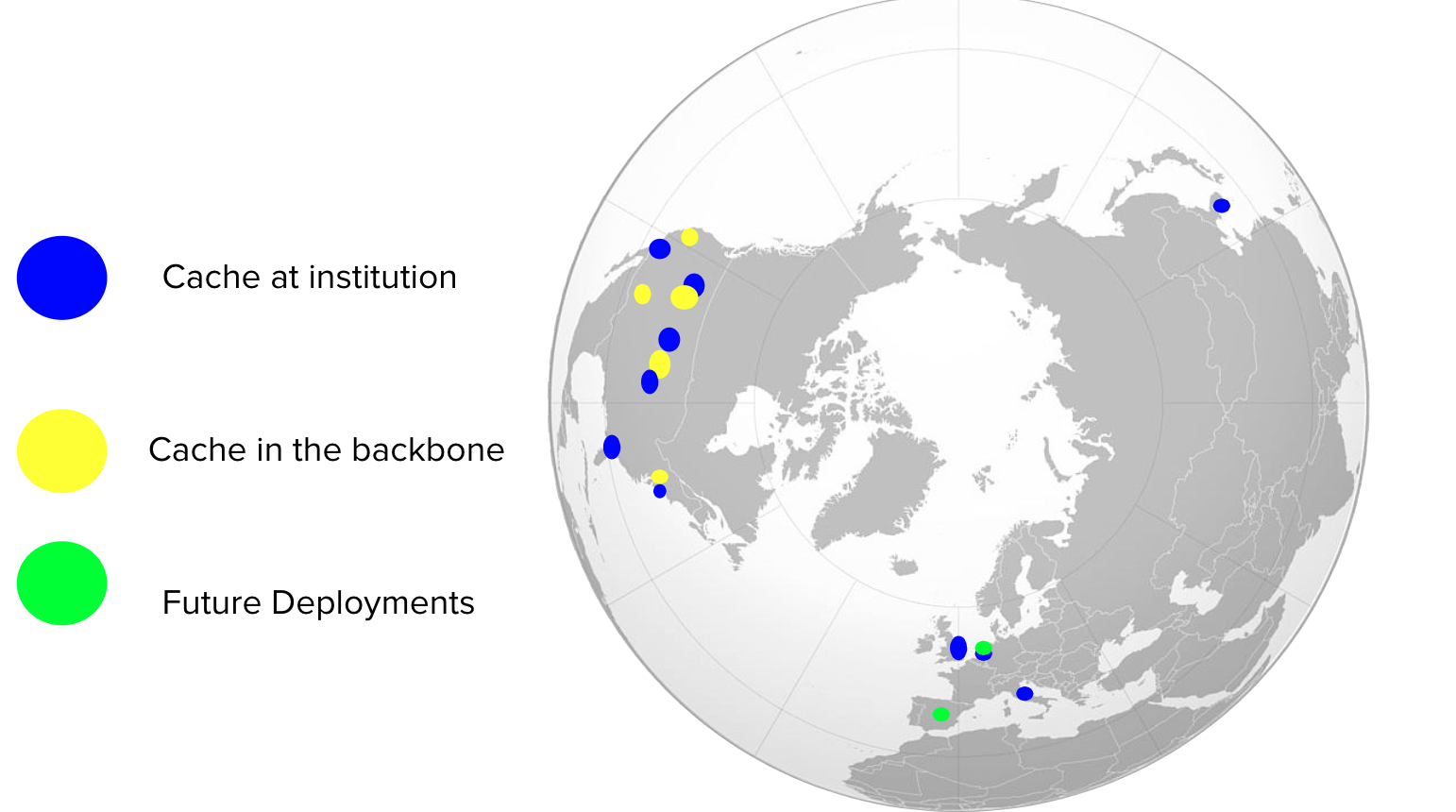}
\caption{Geographical location of the Stashcache infrastructure}
\label{StashcacheWorldide}       
\end{figure}

The OSG through its Virtual Organization (VO) provides several million opportunistic CPU hours for its users and small scientific communities.
The computing resources and the data origins are geographically distributed (Figures \ref{osgCOmputingResources} and \ref{osgDataOrigins}).  Placing caches in the internet backbone reduces the chance that data travels the origin-cache link more than once. Figure \ref{StashcacheUS} show the deployment map of Internet2 and Stashcaches locations in the backbone and at certain academic institutions in the United States. However the International Gravitational Wave Network (IGWN) and the Deep Underground Neutrino Experiment (DUNE) collaborations have their data origins in the US but allocated computing resources abroad hence the need for a worldwide Stashcache infrastructure (Figure \ref{StashcacheWorldide}).

\subsection{Kubernetes deployment}
\label{sec-kubernetes}
Stashcaches are deployed in heterogeneous hardware owned by different institutions. In order to maintain a quality of service and fast turn around from new features to production a Development Ops (DevOps) model is adopted.  This approach allows testing of XRootD features and the ability to roll back to previous versions without the intervention of the system administrators.

In this model Stashcaches are deployed as pods using Kubernetes. A pod consists of an instantiated version of a Docker container. The Stashcache containers are maintained by OSG and are routinely optimized and tested to work with  new XRootD versions. Kubernetes features volume claims and secrets. The first feature bridges the gap  between the stateless micro services that Kubernetes is based on and the stateful XRootD caches; the cache state is the locally stored files. The second feature provides an interface to securely expose X509 certificates to the Stashcache service. 

\section{Stashcache Performance and Future Work}
\label{sec-performance}

Stashcache and CVMFS are combined to create a worldwide content delivery network for science. Table \ref{tab-datasetsize} summarizes the results of data delivered through the Stashcache network. The second column shows the size of the working set for each organization. The working set is defined as the sum of all the unique files accessed in a time frame. The third column shows the total data read by each organization. The fourth one is the division of the third by the second column and represents the number of times the entire working set was read in a time frame. The data reuse factor is an approximation of the cache hit ratio. Future iterations of this work include a file based caching monitoring which provides a more accurate measurement of the cache hit ratio and integration with Scitokens \cite{SciTokens} for authenticated access.

\begin{table}
\centering
\caption{Namespace usage from (September 2019 to February 2020) of the Stashcache infrastructure, data obtained from GRACC \cite{GRACC}.}
\label{tab-datasetsize}       
\begin{tabular}{c | c  | c | c}
\hline
\textbf{Namespace} & Working Set (TB) & Data Read(TB) & Data Reuse Factor\\\hline
 DUNE & 0.014  & 1184  & 84571  \\
 LIGO Public Data & 7.157  & 96 & 75  \\
 Nova & 0.086  & 20 & 233 \\
IGWN & 18.172 & 596  & 33\\\hline 
\end{tabular}
\end{table}

\section*{Acknowledgement}
The authors would like to thank the different funding agencies for this work, in particular the National Science Foundation through the following grants: OAC-1541349, OAC-1826967, OAC-1841530, MPS-1148698. We would also like to thank Internet2 for providing the hardware and support for placing the caches in the backbone.

%
%
\bibliography{XCache.bib}

\end{document}